\documentclass{osa-article}

\journal{osajournal}
\articletype{Research Article}
\begin{document}

\title{Coherent superposition of orthogonal modes}

\author{K. Floettmann,\authormark{1}}

\address{\authormark{1}Deutsches Elektronen-Synchrotron, Notkestra\ss e 85, 22607 Hamburg, Germany}

\email{\authormark{*}Klaus.Floettmann@DESY.De} %% email address is required



\begin{abstract*}
The coherent superposition of orthogonal modes can result in transverse offsets, variations of the Rayleigh length and a reduction of the beam quality factor of the coherent sum of modes in comparison to the incoherent sum. Relations for first and second order moments, the beam quality and the Rayleigh length for the superposition of Hermite-Gauss modes are derived. The Courant-Snyder formalism, which was originally developed in the context of charged particle optics, is applied to propagate an arbitrary coherent sum of orthogonal modes through a lens system. Relations of generating and observable optical functions are highlighted. In the last part of the report the elegant Hermite-Gauss solution is interpreted in terms of generating and observable functions and the solution is decomposed into a sum of standard Hermite-Gauss modes.
\end{abstract*}

%%%%%%%%%%%%%%%%%%%%%%%%%%  body  %%%%%%%%%%%%%%%%%%%%%%%%%%
\section{Introduction}
The treatment of optical and quantum mechanical problems within the framework of the Courant-Snyder theory promises elegant and simplified solutions for many propagation, imaging and matching problems. The Courant-Snyder theory was originally developed in the field of accelerator physics and thus is naturally applicable to classical charged particle optics. It can, however, also be favorably applied to the description of laser modes, because it does not only describe the development of the transverse beam size through linear optical systems, but it also relates the beam size development to the development of the Gouy phase~\cite{Floettmann_2020}. Thus, the field profile of a known mode composition can be determined at each point of an optical system by simple matrix multiplications. Due to the known similarities of the paraxial Helmholtz equation and the Schr\"odinger equation analogue statements hold for a class of quantum mechanical systems.\\
In a broad sense, beams can be formed by ensembles of particles which are moving into a predominant direction, as for example electrons or photons, or by directed wave fields, like electromagnetic waves. A finite intensity and a localization in space, such that an average position and a transverse rms width can be defined at each point of the optical system under consideration, are characteristics of beams, besides their directivity. Based on these simple properties an rms envelope can be defined and the Courant-Snyder theory can be applied.\\
Classical, incoherent beams are widely discussed in terms of their phase space distribution, which is however insufficient for the treatment of coherent or partially coherent beams. K.-J.~Kim proposed the Fourier transform of the cross-spectral density for the brightness definition of synchrotron radiation~\cite{Kim_1986} and noted that his equation resembles the quasi probability distribution which Wigner had introduced in the context of statistical mechanics and which had already been rediscovered by several authors in connection with optical problems. For an expedient review of the Wigner distribution and its relations to quantum mechanics and optics see I.~V.~Bazarov~\cite{Bazarov_2012}. Hereinafter the application of the cross-spectral density or the Wigner distribution for the description and the propagation of partially coherent beams with reference to charged particle optics advanced, especially in the field of synchrotron radiation and FEL physics~\cite{Bazarov_2012, Geloni_2008, Nash_2021}.\\
While the phase space distribution is strictly positive, the Wigner distribution can be locally negative, but it is still positive and normalized in the complete integral over the phase space coordinates. An important property of the Wigner distribution is, that the marginal distributions, i.e., the projections of the distribution onto both phase space coordinates are equal to the marginal distributions of the classical phase space. Due to this identity of the marginal distributions, an rms ellipse can be associated with the Wigner distribution, which is identical to the rms phase space ellipse associated to the phase space density of arbitrary particle distributions~\cite{Floettmann_2003}. Despite the local negativity of the Wigner distribution, it behaves thus with respect to its rms properties just as the classical phase space distribution. It can be mapped through an optical system with the same matrices as the phase space and, just like point like particles move in phase space on concentric ellipses with a phase advance that is described by the Courant-Snyder theory (cf. Fig. 2 in~\cite{Floettmann_2020}), also structures of the Wigner distribution move in the same way on such ellipses.\\
The area of the rms phase space ellipse connects the beam divergence with the beam size and is thus a measure of the beam quality. In charged particle optics the beam quality is hence described by the beam emittance, which is directly proportional to the area of the rms phase space ellipse.
A beam with smaller emittance can be stronger focused and the beam divergence stays smaller than this is the case for a beam with larger emittance. Naturally, the lower limit of the emittance follows Heisenberg's uncertainty principle. A related beam quality factor, the M-square parameter, which is also proportional to the phase space area and thus to the emittance, is employed in laser physics and light optics.\\
The concept of an emittance as conserved quantity of motion, with a lower limit following the Heisenberg relation, and the Courant-Snyder formalism has recently also been applied to describe the manipulation of quantum mechanical vortex particles and the evolution of a wave packet in phase space~\cite{Karlovets_2021}.\\
Despite its solid foundation, the application of the rms envelope and the Courant-Snyder formalism is, however, not in all cases obvious. Especially interference effects, which are negligible in classical accelerator physics, are suspect to lead to deviations from the rms propagation characteristics. This is however not the case as will be discussed below. Also fully or partially coherent beams follow the standard propagation characteristics. However, when describing a beam as coherent sum of basis modes, it will be necessary to clearly distinguish the beam parameters of the coherent sum, which are connected to observable beam sizes, and the parameters of the basis modes, which are not directly observable.\\
In the first part of this paper the effect of the coherent superposition of orthogonal modes will be discussed in detail, and it will be shown how the Courant-Snyder formalism can be used to propagate an arbitrary coherent or incoherent sum of modes through an optical system. In the second part the elegant Hermite-Gauss solution is analyzed with respect to the generating and the observable parameters. It will be shown that it can be described as a superposition of standard Hermite modes and that it describes a beam which follows the Courant-Snyder formalism in the usual way.

\section{Coherent superposition of Hermite-Gauss modes}
It is common practice to describe beams in light optics in terms of the Rayleigh length $Z_R$,  which is the distance from a beam waist over which the transverse beam size increases by a factor square root of 2. In the Courant-Snyder theory the more general ${\beta}$-function is employed. In a free drift the ${\beta}$-function develops as:
\begin{equation}
	\beta (z) = {\beta _0}\left( {1 + \frac{{{z^2}}}{{\beta _0^2}}} \right),
	\label{1.1}
\end{equation}
where $z = 0$ is a focus position. The ${\beta}$-function at the focus, ${\beta_0}$, corresponds to the Rayleigh length.\\
The transverse rms size is then given by
\begin{equation}
	\sigma (z) = \sqrt {\varepsilon \beta (z)},
	\label{1.2}
\end{equation}
where ${\varepsilon}$ denotes the beam emittance.\\
The ${\beta}$-function describes an optical system independent of the specific characteristics of a beam and the emittance connects the transverse rms size and the $\beta$-function. The emittance is related to the beam quality factor ${M^2}$, which is commonly used in laser science, by the relation $\varepsilon  = \frac{{{M^2}}}{{2k}}$, with the wave-number $k = {\raise0.7ex\hbox{${2\pi }$} \!\mathord{\left/ {\vphantom {{2\pi } \lambda }}\right.\kern-\nulldelimiterspace}\!\lower0.7ex\hbox{$\lambda $}}$. $\lambda$ denotes the wavelength of the radiation. For a more thorough discussion of the $\beta$-function and its relation to standard laser parameters see~\cite{Floettmann_2020}.\\ 
As will be discussed below, the $\beta$-function (Rayleigh length) of a coherent sum of modes does not correspond to the $\beta$-function (Rayleigh length) of the individual modes. Moreover, the focus position is found to be shifted by the coherent superposition of the modes. This makes it necessary to distinguish the parameters of the generating modes from the parameters of the coherent sum, which is done by adding an index $g$ to the generating mode parameters, where required.\\
The paraxial wave equation can be solved in the form of a superposition of Hermite-Gaussian modes. The modes constitute a complete and orthogonal basis of solutions. The coherent superposition of modes leads to interference terms in the mathematical description, which are absent when an incoherent superposition is assumed but which are relevant for the beam characteristics of the wave. In the following the influence of the interference terms on the moments of the intensity distribution will be discussed. As usual the transverse position and the size of the wave are described by the first direct and the second central moment of the intensity distribution. The calculations are in general straightforward but lengthy. Assistance by a symbolic computation program is highly appreciated. Only the main results are summarized, while intermediate results are suppressed.\\ 
The moments are calculated for the projections of the two-dimensional transverse distribution onto the axis of the uncoupled coordinate system, which reduces the problem to the 1D case. The transverse coordinate is denoted by $x$, while $z$ denotes the longitudinal direction of predominant motion. A free drift is considered and the focus position -- designated by the index 0 -- corresponds to the origin of the coordinate system.\\
In terms of the generating $\beta$-function ${\beta _g} = {\beta _g}(z)$ Hermit-Gauss modes in a drift are given by:
\begin{equation}
	{E_n} = \frac{1}{{\sqrt {{2^n}n!} }}{\left( {\frac{k}{{\pi {\beta _g}}}} \right)^{1/4}} {H_n}\left( {\sqrt {\frac{k}{{{\beta _g}}}} x} \right){e^{ - \frac{{k{x^2}}}{{2{\beta _g}}}}}{e^{i\frac{1}{2}\left[ {\frac{{k\,{x^2}{z_g}}}{{{\beta _{{g_0}}}\;{\beta _g}}} - \,(2n + 1){\mathop{\rm atan}\nolimits} \left( {\frac{{{z_g}}}{{{\beta _{{g_0}}}}}} \right) + {\varphi _n}} \right]}} ,	
\label{1.3}
\end{equation}
where $n$ is the mode number, ${\varphi _n}$ is an arbitrary phase which subsumes also the term $kz - \omega t$ and $H_n$  is a Hermite polynomial. The amplitude distribution of the field, Eq.~\eqref{1.3}, consists basically of a product of a Hermite polynomial and a Gaussian density distribution.\\
The generating function of the Hermite polynomials is
\begin{equation}
H_n =   {\left( { - 1} \right)^n}\sum\limits_{l + 2m = n} {\frac{{n!}}{{l!m!}}} {\left( { - 1} \right)^{l + m}}(2x)^l.
\label{1.3a}
\end{equation}
Table~\ref{Tab_1} summarizes the first Hermite polynomials for further reference.\\
\begin{table}[hhh]
\centering
\begin{tabular}{l}
\hline
	${H_0}(x) = 1$	\\
	${H_1}(x) = 2x$	\\
	${H_2}(x) = {\left( {2x} \right)^2} - 2$	\\
	${H_3}(x) = {\left( {2x} \right)^3} - 6\left( {2x} \right)$	\\
	${H_4}(x) = {\left( {2x} \right)^4} - 12{\left( {2x} \right)^2} + 12$\\
\hline      
\end{tabular}
\caption{The first five Hermite polynomials.}
\label{Tab_1}
\end{table}
Hermite polynomials are orthogonal with respect to the exponential weight function
\begin{equation}
\int {{e^{ - {x^2}}}} {H_n}(x){H_m}(x)dx = \sqrt \pi  {2^n}n!{\delta _{nm}}.
\label{1.3b}
\end{equation}
Thus, the arguments of all polynomials and exponential amplitude terms in the coherent sum have to be equal to make use of the orthogonality condition. The modes are hence superimposed without relative transverse offset and with the same generating beta function, which implies that all modes reach a focus at the same position.\\
Eq.~\eqref{1.3} is normalized such that the intensity $\int{{E_n}E_n^* dx = 1} $. Here $E_n^* $ is conjugate to ${E_n}$. Integrals span throughout the text from minus to plus infinity. In order to maintain the normalization when two modes with mode numbers $n$ and $m$ ($n \ne m$) are superimposed the relative intensity contributions of the two modes ${a_n}$ and ${a_m}$ need to be normalized such that $a_n^2 + a_m^2 = 1$.\\ 
The integrated intensity of the coherent sum ${a_n}{E_n} + {a_m}{E_m}$ reads then as
\begin{equation}
\begin{gathered}
I = \int {\left[ {{a_n}{E_n} + {a_m}{E_m}} \right]\left[ {{a_n}E_n^* + {a_m}E_m^*} \right]dx} \hfill  \\
 = \int {\left[ {a_n^2{E_n}E_n^* + {a_n}{a_m}\left( {{E_n}E_m^* + {E_m}E_n^*} \right) + a_m^2{E_m}E_m^*} \right]dx}\hfill  \\
 = \int {a_n^2{E_n}E_n^*dx + \int {a_m^2{E_m}E_m^*dx} }. \hfill 
\end{gathered}
\label{1.4}
\end{equation}
The condition $\int {{E_n}E_m^*dx}  = \int {{E_m}E_n^*dx = 0}$ follows from the orthogonality of the modes and ensures energy conservation. Note, that ${\mathop{\Re}\nolimits} \left( {{E_n}E_m^*} \right) = {\mathop{\Re}\nolimits} \left( {E_n^*{E_m}} \right)$ and ${\mathop{\Im}\nolimits} \left( {{E_n}E_m^*} \right) =  - {\mathop{\Im}\nolimits} \left( {E_n^*{E_m}} \right)$, so that the sum of both terms is real.\\
The orthogonality of the modes leads also to the condition that most combinations of the mode numbers $n$ and $m$ don't result in a contribution to the first and second moment of the field distribution. For the transverse position $\bar x = \int {\left[ {{a_n}{E_n} + {a_m}{E_m}} \right]\left[ {{a_n}E_n^* + {a_m}E_m^*} \right]x\,dx} $ the interference term $\int {{a_n}{a_m}\left( {{E_n}E_m^* + {E_m}E_n^*} \right)x\,dx} $ is zero in all cases, except for $m = n + 1$. This is explained by the fact that $x{H_n}$ contains the same polynomial orders as ${H_{n + 1}}$ (cf. Table~\ref{Tab_1}). The mathematical structure of the integrands is thus similar to the square of a single mode and a kind of modified orthogonality condition is realized. Equally the interference term in the second moment is zero for all cases except for $m = n + 2$. Again, ${x^2}{H_n}$ contains the same polynomial orders as ${H_{n + 2}}$ (cf. Table~\ref{Tab_1}) and the integrands are hence nonzero. The conditions read as
\begin{equation} 
\int {{e^{ - {x^2}}}} {H_n}(x){H_m}(x)xdx  = \;\left\{ \begin{gathered}
 \sqrt \pi {2^n}n!\hspace{3pt} (n+1)\hspace{8pt}{\text{for}}\hspace{8pt} m=n+1 \hfill \\
  \hspace{30pt}0\hspace{35pt}{\text{else}} \hfill \\ 
\end{gathered}  \right.
\label{Eq.1.4a}
\end{equation}
and
\begin{equation} 
\int {{e^{ - {x^2}}}} {H_n}(x){H_m}(x)x^2dx  = \;\left\{ \begin{gathered}
 \sqrt \pi {2^n}n!\hspace{3pt} (n+1)(n+2)\hspace{8pt}{\text{for}}\hspace{8pt} m=n+2 \hfill \\
  \hspace{40pt}0\hspace{53pt}{\text{else}}. \hfill \\ 
\end{gathered}  \right.
\label{Eq.1.4b}
\end{equation}  
 Clearly more mode combinations contribute to higher order moments, which are however beyond the scope of this paper.\\
In the following the influence of mode combinations of the case $m = n + 1$  and of the $m = n + 2$ case are discussed separately before a generalization to an arbitrary combination of modes is presented. The relative mode intensities ${a_n}$ are assumed to be normalized as $\sum\limits_{n = 0}^\infty  {a_n^2}  = 1$. As abbreviations 
\begin{equation}
\begin{array}{l}
{S_{\text{in}}} = \sum\limits_{n = 0}^\infty  {a_n^2} \left( {2n + 1} \right)\\
{S_1} = 2\sum\limits_{n = 0}^\infty  {{a_n}{a_{n + 1}}\sqrt {\left( {n + 1} \right)} } \\
{S_2} = 2\sum\limits_{n = 0}^\infty  {{a_n}{a_{n + 2}}\sqrt {\left( {n + 1} \right)\left( {n + 2} \right)} } 
\end{array}
\label{1.5}
\end{equation}
are introduced, where ${S_{\text{in}}}$ describes the incoherent part of the relations, while ${S_1}$ and ${S_2}$ correspond to the contributions of the $m = n + 1$ and the $m = n + 2$ case, respectively. The sums are positive and  ${S_{\text{in}}}$ is larger than ${S_1}$ or ${S_2}$. 

\section{The case $m=n+1, S_2=0$}
\begin{figure}[htb]
\includegraphics[width=\textwidth]{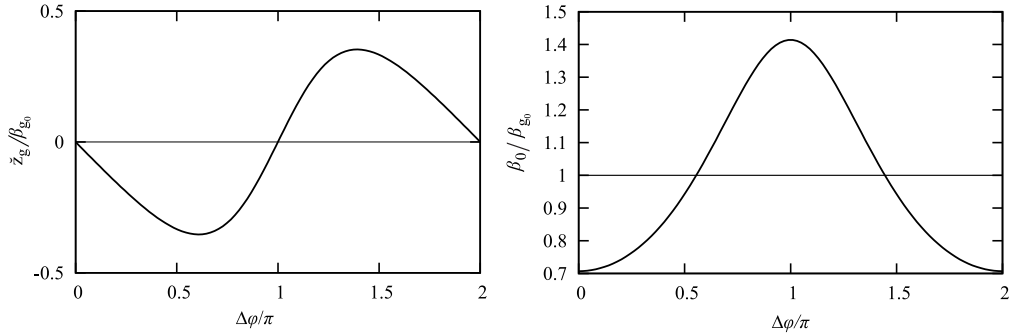}
\caption{Focus position ${z_g}/{\beta _{{g_0}}}$ (left), and ${\beta _0}/{\beta _{{g_0}}}$ (right) for the case $m = n + 1$. Two modes with equal intensity are assumed.}
\label{fig.1}
\end{figure}

The first moment of coherently superimposed modes is given as
\begin{equation}
\begin{array}{l}
\bar x = \int {\sum\limits_{n = 0}^\infty  {{a_n}{E_n}} } \sum\limits_{n = 0}^\infty  {{a_n}E_n^*} x\;dx\\
 = {S_1}\sqrt {\frac{{{\beta _{{g_0}}}}}{{2k}}} \left[ {\cos \left( {\frac{{\Delta {\varphi _1}}}{2}} \right) - \sin \left( {\frac{{\Delta {\varphi _1}}}{2}} \right)\frac{{{z_g}}}{{{\beta _g}}}} \right],
\end{array}
\label{1.6}
\end{equation}
where $\Delta {\varphi _1} = {\varphi _n} - {\varphi _{n + 1}}$ denotes the phase difference between the first and the subsequent mode. Here it is assumed that the phase difference for all mode combinations in the sum is the same, which is not necessarily the case, but leads to simplified equations. Coherence requires also that the phase difference is constant over a sufficient time interval, which is only possible if both modes have the same frequency as ${\varphi _n}$  subsumes the term $kz - \omega t$. Note, that at no phase difference offset and angle are simultaneously zero.\\
While Eq.~\eqref{1.6} leaves the direct second moment unchanged, the central second moment
\begin{equation}
	\left\langle {{x^2}} \right\rangle  = \int {\sum\limits_{n = 0}^\infty  {{a_n}{E_n}} } \sum\limits_{n = 0}^\infty  {{a_n}E_n^*} {x^2}\;dx - {\bar x^2}
\label{1.7}
\end{equation}
and thus, the transverse rms size $\sigma  = \sqrt {\left\langle {{x^2}} \right\rangle } $ is modified.\\
The condition $m = n + 1$ leads for the first term in Eq.~\eqref{1.7} to
\begin{equation}
\int {\sum\limits_{n = 0}^\infty  {{a_n}{E_n}} } \sum\limits_{n = 0}^\infty  {{a_n}E_n^*} {x^2}\;dx = {S_{\text{in}}}\frac{{{\beta _{{g_0}}}}}{{2k}}\left( {1 + \frac{{z_g^2}}{{\beta _{{g_0}}^2}}} \right).
\label{1.8}
\end{equation}
This term describes the incoherent addition and thus the mode interference does not contribute to it for $m = n + 1$. However, the beam size follows as:
\begin{equation}
{\sigma ^2} = \frac{{{\beta _{{g_0}}}}}{{2k}}\left\{ {{S_{\text{in}}}\left( {1 + \frac{{z_g^2}}{{\beta _{{g_0}}^2}}} \right) - S_1^2{{\left[ {\cos \left( {\frac{{\Delta {\varphi _1}}}{2}} \right) - \sin \left( {\frac{{\Delta {\varphi _1}}}{2}} \right)\frac{{{z_g}}}{{{\beta _{{g_0}}}}}} \right]}^2}} \right\}.
\label{1.9}
\end{equation}
The square over the bracket of the second term leads to a linear position dependence. Thus, the beam size minimum is not reached ${z_g} = 0$, or, in other words, the phase front of the coherent sum is not straight at ${z_g} = 0$.\\
Instead the beam size minimum is reached at
\begin{equation}
{\mathord{\buildrel{\lower3pt\hbox{$\scriptscriptstyle\smile$}} 
\over z} _g} =  - {\beta _{{g_0}}}\frac{{S_1^2\sin \Delta {\varphi _1}}}{{2\left[ {{S_{\text{in}}} - S_1^2{{\sin }^2}\left( {\frac{{\Delta {\varphi _1}}}{2}} \right)} \right]}}.
\label{1.10}
\end{equation} 
Rewriting Eq.~\eqref{1.9} in terms of the shifted focus position $z = {z_g} - {\mathord{\buildrel{\lower3pt\hbox{$\scriptscriptstyle\smile$}} 
\over z} _g}$ leads to
\begin{equation}
{\sigma ^2} = \frac{{{\beta _{{g_0}}}}}{{2k}}\left\{ {\frac{{{S_{\text{in}}}({S_{\text{in}}} - S_1^2)}}{{{S_{\text{in}}} - S_1^2{{\sin }^2}\left( {\frac{{\Delta {\varphi _1}}}{2}} \right)}} + \left[ {{S_{\text{in}}} - S_1^2{{\sin }^2}\left( {\frac{{\Delta {\varphi _1}}}{2}} \right)} \right]\frac{{z^2}}{{\beta _{{g_0}}^2}}} \right\},
\label{1.11}
\end{equation}
which is symmetric with respect to the position $z = 0$, but still not in the standard form Eq.~\eqref{1.1}. Thus the $\beta$-function of the coherent sum of the modes differs from the generating $\beta$-function ${\beta _g}$. Or, in other words, the Rayleigh length of the coherent sum of the modes differs from the Rayleigh length of the individual modes.
Eq.~\eqref{1.11} has the form ${\sigma ^2} = \frac{{{\beta _{{g_0}}}}}{{2k}}\left( {A + B\frac{{{z^2}}}{{\beta _{{g_0}}^2}}} \right)$ with the parameters
\begin{equation}
\begin{array}{l}
A = \frac{{{S_{\text{in}}}({S_{\text{in}}} - S_1^2)}}{{{S_{\text{in}}} - S_1^2{{\sin }^2}\left( {\frac{{\Delta {\varphi _1}}}{2}} \right)}} \\
B = {S_{\text{in}}} - S_1^2{\sin ^2}\left( {\frac{{\Delta {\varphi _1}}}{2}} \right).
\end{array}
\label{1.12}
\end{equation}
While the transverse rms size at the focus is proportional to $\sqrt A $, the far field diffraction angle is proportional to $\sqrt B $, which leads to the relations:
\begin{equation}
\begin{array}{l}
{\beta _0} = {\beta _{{g_0}}}\sqrt {\frac{A}{B}} \\
= {\beta _{{g_0}}}\frac{{\sqrt {{S_{\text{in}}}({S_{\text{in}}} - S_1^2)} }}{{{S_{\text{in}}} - S_1^2{{\sin }^2}\left( {\frac{{\Delta {\varphi _1}}}{2}} \right)}}
\end{array}
\label{1.13}
\end{equation}
and
\begin{equation}
\begin{array}{l}
\varepsilon  = \frac{1}{{2k}}\sqrt {AB}\\
= \frac{1}{{2k}}\sqrt {{S_{\text{in}}}({S_{\text{in}}} - S_1^2)}.
\end{array}
\label{1.14}
\end{equation}
With these relations Eq.~\eqref{1.11} is transformed into the standard form ${\sigma ^2} = \varepsilon {\beta _0}\left\{ {1 + \frac{{{z^2}}}{{\beta _0^2}}} \right\}$.\\
The interference term in the superposition of two modes with $m = n + 1$ leads thus to a phase dependent transverse offset, but also to a shift of the focus position, and a variation of the transverse size, the emittance and the $\beta$-function.\\
A simple example is the addition of only two modes. The interference term gets maximal when both modes contribute with the same intensity, the relevant equations reduce then to
\begin{equation}
\bar x = \sqrt {\frac{{n + 1}}{{2k}}{\beta _{{g_0}}}} \left[ {\cos \left( {\frac{{\Delta {\varphi _1}}}{2}} \right) - \sin \left( {\frac{{\Delta {\varphi _1}}}{2}} \right)\frac{{{z_g}}}{{{\beta _g}}}} \right]
\label{1.15}
\end{equation}
\begin{equation}
{\mathord{\buildrel{\lower3pt\hbox{$\scriptscriptstyle\smile$}} 
\over z} _g} =  - {\beta _{{g_0}}}\frac{{\sin \Delta {\varphi _1}}}{{3 + \cos \left( {\Delta {\varphi _1}} \right)}}
\label{1.16}
\end{equation}
\begin{equation}
	{\beta _0} = {\beta _{{g_0}}}\frac{{2\sqrt 2 }}{{3 + \cos \left( {\Delta {\varphi _1}} \right)}}
\label{1.17}
\end{equation}
and
\begin{equation} 
	\varepsilon  = \frac{{\sqrt 2 }}{{2k}}n + 1.
\label{1.18}
\end{equation}
Figure~\ref{fig.1} shows the phase dependence of the focus position and of the minimal $\beta$-function.\\ 

The shift of the focus position [Eq.~\eqref{1.10}] is on the order of the generating $\beta$-function at the focus, i.e., on the order of the Rayleigh length and the $\beta$-function varies between $\sqrt 2 $ and ${\raise0.7ex\hbox{$1$} \!\mathord{\left/{\vphantom {1 {\sqrt 2 }}}\right.\kern-\nulldelimiterspace}\!\lower0.7ex\hbox{${\sqrt 2 }$}}$ times the generating $\beta$-function.\\
The emittance should be compared to the emittance of the incoherent addition, which is determined solely by ${S_{\text{in}}}$ and thus yields ${\varepsilon _{\text{in}}} = \frac{n+1}{k}$ for the case under consideration. The coherent addition leads hence to an emittance reduction by a factor ${\raise0.7ex\hbox{$1$} \!\mathord{\left/ {\vphantom {1 {\sqrt 2 }}}\right.\kern-\nulldelimiterspace}\!\lower0.7ex\hbox{${\sqrt 2 }$}}$.\\
Finally the factor $\sqrt {\frac{{n + 1}}{{2k}}{\beta _{{g_0}}}} $ can be approximated by ${\sigma}=\sqrt {\varepsilon {\beta _0}} $ in Eq.~\eqref{1.15} to see that the transverse offset near the focus can become roughly as large as the rms beam size.

\section{The case $m=n+2, S_1=0$}
Another mode combination which influences the second order moment is the case $m = n + 2$. While in the previous case the offset, i.e., the second term of Eq.~\eqref{1.7} was not zero, the offset is zero for $m = n + 2$, but the first term of Eq.~\eqref{1.7} is modified.
The transverse rms size is given in this case as:
\begin{equation}
{\sigma ^2} = \frac{{{\beta _{{g_0}}}}}{{2k}}\left\{ {{S_{\text{in}}}\left( {1 + \frac{{z_g^2}}{{\beta _{{g_0}}^2}}} \right) + {S_2}\left[ {\cos \left( {\frac{{\Delta {\varphi _2}}}{2}} \right)\left( {1 - \frac{{z_g^2}}{{\beta _{{g_0}}^2}}} \right) + 2\sin \left( {\frac{{\Delta {\varphi _2}}}{2}} \right)\frac{{{z_g}}}{{{\beta _{{g_0}}}}}} \right]} \right\},
\label{1.19}
\end{equation}
where $\Delta {\varphi _2}$ denotes the phase difference between the two modes.\\
Again, the focus is shifted due to the linear term in ${z_g}$. The minimum transverse size is reached at:
\begin{equation}
{\mathord{\buildrel{\lower3pt\hbox{$\scriptscriptstyle\smile$}}\over z} _g} = {\beta _{{g_0}}}\frac{{{S_2}\sin \left( {\frac{{\Delta {\varphi _2}}}{2}} \right)}}{{{S_{\text{in}}} - {S_2}\cos \left( {\frac{{\Delta {\varphi _2}}}{2}} \right)}}.
\label{1.20}
\end{equation}
Introducing $z = {z_g} - {\mathord{\buildrel{\lower3pt\hbox{$\scriptscriptstyle\smile$}}\over z} _g}$ into Eq.~\eqref{1.19} leads to ${\sigma ^2} = \frac{{{\beta _{{g_0}}}}}{{2k}}\left( {A + B\frac{{{z^2}}}{{\beta _{{g_0}}^2}}} \right)$ with
\begin{equation}
\begin{array}{l}
A = \frac{{S_{\text{in}}^2 - S_2^2}}{{{S_{\text{in}}} - {S_2}\cos \left( {\frac{{\Delta {\varphi _2}}}{2}} \right)}}\\
B = {S_{\text{in}}} - {S_2}\cos \left( {\frac{{\Delta {\varphi _2}}}{2}} \right)
\end{array}
\label{1.21}
\end{equation}
and thus to
\begin{equation}
{\beta _0} = {\beta _{{g_0}}}\frac{{\sqrt {S_{\text{in}}^2 - S_2^2} }}{{{S_{\text{in}}} - {S_2}\cos \left( {\frac{{\Delta {\varphi _2}}}{2}} \right)}}
\label{1.22}
\end{equation}
\begin{equation}
	\varepsilon  = \frac{1}{{2k}}\sqrt {S_{\text{in}}^2 - S_2^2}.
\label{1.23}
\end{equation}
The emittance is reduced in comparison to the incoherent addition, as is immediately visible from Eq.~\eqref{1.23}. The effects on the focus position and on the $\beta$-function for the simple example of two modes with equal intensity are of similar magnitude as discussed above for the $m=n+1$ case. 
 
\section{The general case}
The generalization is now straight forward and follows the steps outlined above. In order to simplify the equations, the phase difference $\Delta {\varphi _1}$ between two successive modes and $\Delta {\varphi _2}$ between one and the next but one mode where introduced above, assuming already that the phase difference is the same for all relevant mode combinations in the sums for the different cases. In the general case, i.e., when arbitrary modes are superimposed, this requires that $\Delta {\varphi _2} = 2\Delta \varphi {}_1$.\\ 
With this assumption the focus shift,  $\beta$-function and emittance are found as
\begin{equation}
	{\mathord{\buildrel{\lower3pt\hbox{$\scriptscriptstyle\smile$}} 
\over z} _g} = {\beta _{{g_0}}}\frac{{\left( {2{S_2} - S_1^2} \right)\sin \Delta {\varphi _1}}}{{2\left[ {{S_{\text{in}}} - {S_2}\cos \Delta {\varphi _1} - S_1^2{{\sin }^2}\left( {\frac{{\Delta {\varphi _1}}}{2}} \right)} \right]}}
\label{1.24}
\end{equation}
\begin{equation}
{\beta _0} = {\beta _{{g_0}}}\frac{{\sqrt {S_{\text{in}}^2 - {S_{\text{in}}}S_1^2 - S_2^2 + {S_2}S_1^2} }}{{{S_{\text{in}}} - {S_2}\cos \Delta {\varphi _1} - S_1^2{{\sin }^2}\left( {\frac{{\Delta {\varphi _1}}}{2}} \right)}}
\label{1.25}
\end{equation}
\begin{equation}
\varepsilon  = \frac{1}{{2k}}\sqrt {S_{\text{in}}^2 - {S_{\text{in}}}S_1^2 - S_2^2 + {S_2}S_1^2}.
\label{1.26}
\end{equation}

\begin{figure}[hhh!!!]
\includegraphics[width=\textwidth]{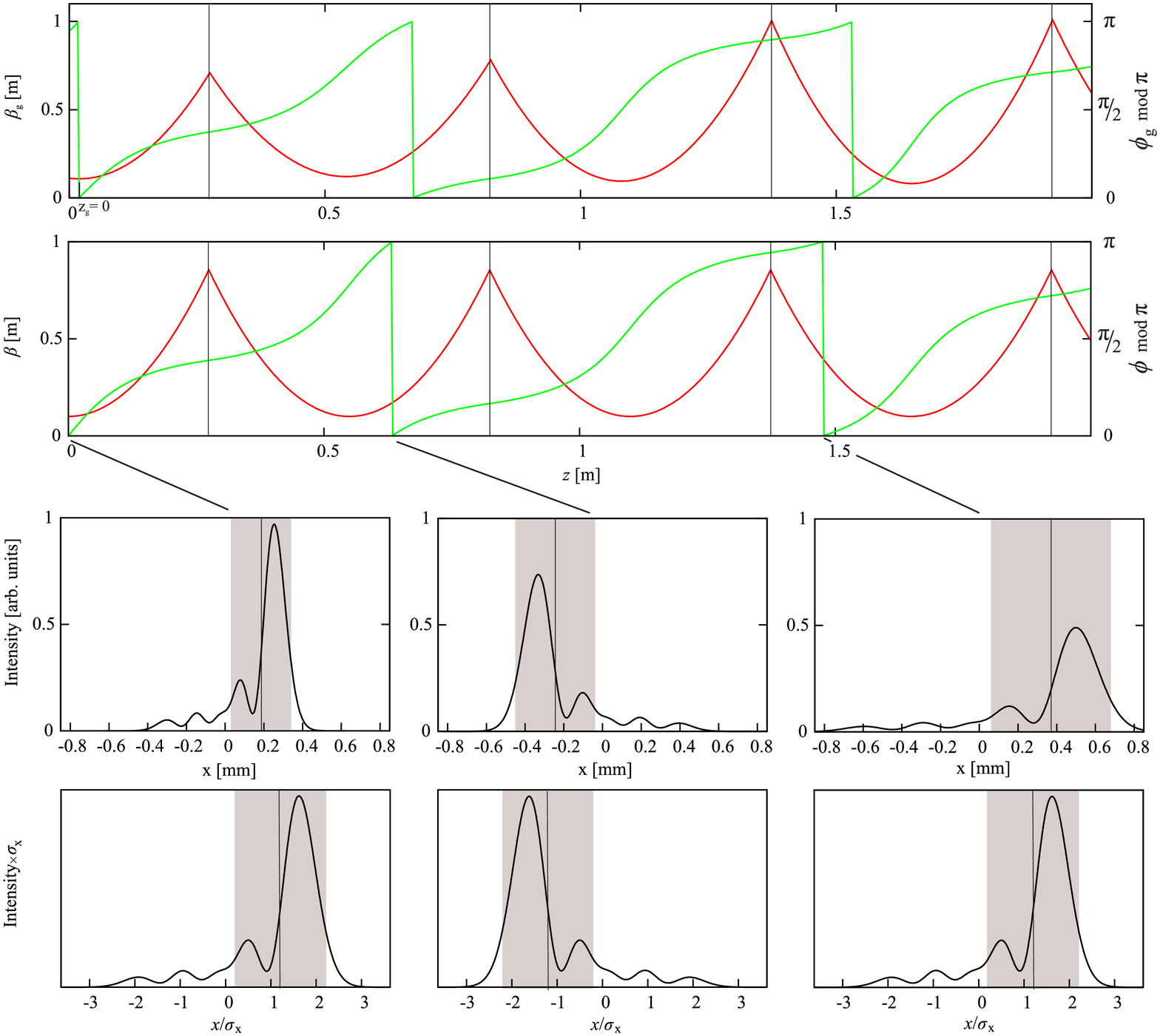}
\caption{Propagation of a coherent sum of modes through a periodic optical system. Lens positions are indicated by vertical lines in the upper two panels. The top panel shows the development of the generating $\beta$-function (red, left scale) and the generating phase advance modulo $\pi $ (green, right scale). The middle panel shows the corresponding observable $\beta$-function (red, left scale) and phase advance modulo $\pi $ (green, right scale). The bottom rows show the transverse intensity distribution at three locations with a phase advance of $\pi $ with scaled and unscaled coordinates. Grey shaded areas indicate the plus/minus one sigma beam width around the average position.}
\label{fig.2}
\end{figure}

Figure 2 shows as example the propagation of a coherent sum of modes through a periodical optical system. The calculations follow the standard procedures of the Courant-Snyder formalism. The only difference is, that two sets of optical functions are traced trough the system, i.e., one set of generating functions and one set of observable functions. In the incoherent case these two sets are identical.\\
An initial $\beta$-function of ${\beta _0}$=10 cm and a phase advance of 140$^\circ$ between two lenses are chosen. With these parameters the optical system is fully determined. The calculation of the optical functions $\alpha $, $\beta $ and $\gamma $ is derived from the abcd-matrix of drift and thin lens elements in the standard form and the phase advance is given, as usual, by $\phi  = \int {\frac{1}{\beta }} dz$. For details see~\cite{Floettmann_2020}. The second panel of Figure 2 displays the development of the $\beta$-function and the phase advance modulo $\pi $.\\
The beam in this illustrating example is generated by a superposition of three consecutive modes, $n = 2, 3, 4$. Each mode contributes with the same intensity. The curvature of the phase front of the Hermite-Gauss modes is expressed by the generating $\alpha$-function and the Gouy phase is expressed by the generating phase advance as described in~\cite{Floettmann_2020}. A wavelength of 800~nm is assumed and the first modes starts with $\varphi  = 0$ at ${z_g} = 0$. The other modes are shifted by $ - 0.4\pi $ and $ - 0.8\pi $ relative to the first mode. The start parameter of the generating $\beta$-function is determined by Eq.~\eqref{1.25}. At the focus position of the generating functions ${z_g} = 0$ the standard relations ${\alpha _g} = 0$ and ${\gamma _g} = {\raise0.7ex\hbox{$1$} \!\mathord{\left/{\vphantom {1 {{\beta _g}}}}\right.\kern-\nulldelimiterspace}
\!\lower0.7ex\hbox{${{\beta _g}}$}}$ hold. These parameters are in a first step traced to the position $z = 0$, so that both sets of optical functions refer to the same position. The upper panel of Figure 2 shows the development of the generating $\beta$-function and the generating phase advance which corresponds to the phase of the first mode. While the observable $\beta$-function is periodic, the generating $\beta$-function is not periodic, i.e., while the observable $\beta$-function is properly matched the generating function is not matched.\\
Only the generating functions are used to track the modes through the system, but the observable $\beta$-function describes the rms beam size and the corresponding phase advance determines the imaging condition as demonstrated in the lower two rows of Figure 2. The initial transverse intensity distribution is imaged whenever the phase advance is a multiple of $180^\circ $. Thus, the transverse intensity distribution has the same shape and the transverse offset is, relative to the beam size, the same. For better comparison the plots of the intensity distribution are reproduced with scaled coordinates in the lowest row.\\

\section{The Elegant Hermite-Gauss functions}
Siegman~\cite{Siegman_1973} established the so-called elegant Hermite-Gauss functions as a symmetrized solution of the paraxial Helmholtz equation by introducing a complex argument into the polynomial part of the Hermite-Gauss field description. Elegant Hermite-Gauss solutions are useful to treat several theoretical problems, and are hence a relevant example for a generalized solution of the paraxial Hemholtz equation. Besides in depth studies of more mathematical properties, e.g.~\cite{Zauderer_1986, Wuensche_1989}, also propagation properties of the elegant Hermite-Gauss solutions have been studied~\cite{Saghafi_1998, Saghafi_1998_2, Lue_2000}.\\ 
Beams described by the elegant Hermite-Gauss solution change their transverse shape as they propagate and thus they are not simple modes. They also don't form an orthogonal basis with respect to the transverse coordinate, but rather a biorthogonal set of functions with a corresponding conjugate set.\\
While being mathematically elegant, the interpretation of the complex solution in terms of physical quantities is not straight forward. In the following the elegant solution will be discussed in terms of the generating and the observable $\beta$-function of a coherent sum of modes.\\ Furthermore, the decomposition of the elegant solution is presented.\\
The elegant solution, indicated by the tilde, reads in terms of the generating $\beta$-function as
\begin{equation}
{\tilde E_n} = \sqrt {\frac{2^n n!}{\left( 2n \right)! }} {\left( {\frac{k}{{{\pi \beta _g}}}} \right)^{1/4}}{\left( {\frac{1}{{1 + \frac{{{z_g^2}}}{{\beta _{g_0}^2}}}}} \right)^{n/4}}{H_n}\left( {\sqrt {\frac{k}{{2\left( {{\beta _{g_0}} + i\;{z_g}} \right)}}} x} \right){e^{\left[ {{ - \frac{{k\;{x^2}}}{{2\left( {{\beta _{g_0}} + i\;{z_g}} \right)}} - i\left( {\frac{{n + 1}}{2}} \right)\arctan \left( {\frac{{{z_g}}}{{{\beta _{{g_0}}}}}} \right)}} \right]}}.
\label{1.27}
\end{equation}
Eq.~\eqref{1.27} is normalized analog to the Hermite-Gauss modes, i.e., $\int  {{\tilde E_n} {\tilde E_n^*} dx = 1} $. The first order moment of the elegant solution is zero. Calculating the rms beam size yields:
\begin{equation}
\begin{array}{l}
{\sigma ^2} = \frac{{{\beta _{g_0}}}}{{2k}}\left( {A + B\frac{{{z^2}}}{{\beta _{{g_0}}^2}}} \right)\\
A = \frac{{4n - 1}}{{2n - 1}}\\
B = 2n + 1.
\end{array}
\label{1.28}
\end{equation}
Which leads to
\begin{equation}
{\beta _0} = {\beta _{{g_0}}}\sqrt {\frac{{4n - 1}}{{4{n^2} - 1}}} 
\label{1.29}
\end{equation}
\begin{equation}
\varepsilon  = \frac{1}{{2k}}\sqrt {\frac{{\left( {4n - 1} \right)\left( {2n + 1} \right)}}{{2n - 1}}}. 
\label{1.30}
\end{equation}
The focus position is not shifted, i.e. ${z_g} = z$.\\ 
Other than in the standard Hermite-Gauss solution, where the $\beta$-function at the focus, i.e., the Rayleigh length, is independent of the mode number, the Rayleigh length of the elegant solution scales inversely to $\sqrt{n}$ for $n> 1$, while the transverse beam size at the focus stays nearly constant. Since the ${\beta}$-function describes an optical system independent of the specific characteristics of a beam, one may say that the optics is not fixed in case of the elegant solution. As shown above, this is the result of a coherent superposition of basis modes and a specific property which needs to be taken into account when discussing the propagation of generalized solutions.\\
As example for the decomposition of the elegant solution the case $n=2$ will be explicitly executed below. Since the elegant solution exhibits no offset it is to be expected that the case $m=n+1$ doesn't appear in the sum of orthogonal modes. Moreover, the focus is not shifted, while the observable $\beta$-function at the focus is reduced in comparison to the generating $\beta$-function (cf. Eq.~\eqref{1.29}). These conditions are reached at a phase difference of the modes of $\Delta {\varphi _2} = 2\pi $ (cf. Eqs.~\eqref{1.20} and \ref{1.22}). Since the phase enters with a factor $1/2$ in the exponential (cf. Eq.~\eqref{1.3}) this corresponds to a change of sign.

\section{Decomposition of the elegant solution}
To simplify the notation, the normalization terms ${I_n} = \frac{1}{{\sqrt {{2^n}n!}}}$ for the Hermite-Gauss mode and ${\tilde I_n} = \sqrt {\frac{2^n n!}{\left( 2n \right)! }} $ for the elegant solution are introduced, and a common factor ${\left( {\frac{k}{{{\pi \beta _g}}}} \right)^{1/4}}{e^{ - \frac{{k{x^2}}}{{2{\beta _g}}}}}{e^{i\frac{1}{2}\left[ {\frac{{kz}}{{{\beta _{{g_0}}}{\beta _g}}}} \right]}}$ is dropped. This factor has to be included in the final equations.\\
The elegant solution (Eq.~\eqref{1.27}) now has the form:
\begin{equation}
{\tilde E_n} = {\tilde I_n}{\left( {\frac{1}{{1 + \frac{{{z^2}}}{{\beta _{g_0}^2}}}}} \right)^{n/4}}{H_n}\left( {\sqrt {\frac{k}{{2\left( {{\beta _{g_0}} + i\;z} \right)}}} x} \right){e^{i\frac{1}{2}\left[ { - (n + 1){\rm atan}\left( {\frac{z}{{{\beta _{{g_0}}}}}} \right)} \right]}},
\label{1.31}
\end{equation}
while the Hermite-Gauss mode (Eq.~\eqref{1.3}) reads as:
\begin{equation}
{E_n} = {I_n}{H_n}\left( {\sqrt {\frac{k}{{{\beta _g}}}} x} \right){e^{i\frac{1}{2}\left[ { - \,(2n + 1){\mathop{\rm atan}\nolimits} \left( {\frac{z}{{{\beta _g}}}} \right) + {\varphi _n}} \right]}},
\label{1.32}
\end{equation}
where already ${z_g} = z$ is used.\\
Introducing ${H_2} = 4{x^2} - 2$ into the elegant solution leads to
\begin{equation}
{\tilde E_2} = {\tilde I_2}{\left( {\frac{1}{{1 + \frac{{{z^2}}}{{\beta _{g_0}^2}}}}} \right)^{1/2}}\left( {\frac{{2k}}{{\left( {{\beta _{g_0}} + i\;z} \right)}}{x^2} - 2} \right){e^{ - i\frac{3}{2}\left( {\frac{z}{{{\beta _{{g_0}}}}}} \right)}},
\label{1.33}
\end{equation}
which is transformed with the relation $\frac{1}{{1 + i\frac{z}{{{\beta _{g_0}}}}}} = \frac{1}{{\sqrt {1 + \frac{{{z^2}}}{{\beta _{g_0}^2}}} }}{e^{ - i{\mathop{\rm atan}\nolimits} \left( {\frac{z}{{{\beta _{g_0}}}}} \right)}}$ into:
\begin{equation}
{\tilde E_2} = {\tilde I_2}\frac{1}{2}\left[ {4\frac{k}{{{\beta _g}}}{x^2} - 2} \right]{e^{ - i\frac{5}{2}{\mathop{\rm atan}\nolimits} \left( {\frac{z}{{{\beta _{{g_0}}}}}} \right)}} + {\tilde I_2}\left( {{e^{ - i{\mathop{\rm atan}\nolimits} \left( {\frac{z}{{{\beta _{{g_0}}}}}} \right)}} - \frac{2}{{\sqrt {1 + \frac{{{z^2}}}{{\beta _{g_0}^2}}} }}} \right){e^{ - i\frac{3}{2}{\mathop{\rm atan}\nolimits} \left( {\frac{z}{{{\beta _{{g_0}}}}}} \right)}},
\label{1.34}
\end{equation}
where ${e^{ - i{\mathop{\rm atan}\nolimits} \left( {\frac{z}{{{\beta _{{g_0}}}}}} \right)}} - {e^{ - i{\mathop{\rm atan}\nolimits} \left( {\frac{z}{{{\beta _{{g_0}}}}}} \right)}} = 0$ has been added.\\ 
The first term can now be replaced by $\frac{1}{2}\frac{{{{\tilde I}_2}}}{{{I_2}}}{E_2}$, while in the second term ${e^{ - i{\mathop{\rm atan}\nolimits} \left( {\frac{z}{{{\beta _{{g_0}}}}}} \right)}}$ is replaced by $ \frac{{1 - i\left( {\frac{z}{{{\beta _{{g_0}}}}}} \right)}}{{\sqrt {1 + \frac{{{z^2}}}{{\beta _{{g_0}}^2}}} }}$
\begin{equation}
{\tilde E_2} = \frac{1}{2}\left( {\frac{{{{\tilde I}_2}}}{{{I_2}}}{E_2} - 2{{\tilde I}_2}\frac{{1 + i\left( {\frac{z}{{{\beta _g}}}} \right)}}{{\sqrt {1 + \frac{{{z^2}}}{{\beta _{{g_0}}^2}}} }}{e^{ - i\frac{3}{2}{\mathop{\rm atan}\nolimits} \left( {\frac{z}{{{\beta _{{g_0}}}}}} \right)}}} \right),
\label{1.35}
\end{equation}
which leads with $ 1 + i\frac{z}{{{\beta _{{g_0}}}}} = \sqrt {1 + \frac{{{z^2}}}{{{\beta _{{g_0}}}}}} {e^{i{\mathop{\rm atan}\nolimits} \left( {\frac{z}{{{\beta _{{g_0}}}}}} \right)}}$ to 
\begin{equation}
{\tilde E_2} = \frac{1}{2}\left( {\frac{{{{\tilde I}_2}}}{{{I_2}}}{E_2} - 2\frac{{{{\tilde I}_2}}}{{{I_0}}}{E_0}} \right).
\label{1.36}
\end{equation}
The negative sign corresponds, as already mentioned, to a phase difference of $2\pi $. In the following the sign will be kept however.\\
The calculation of other orders works in the same way, but gets increasingly complex with increasing order. The results for the first five solutions are summarized as:
\begin{equation}
\begin{array}{l}
{{\tilde E}_0} = {E_0}\\
{{\tilde E}_1} = 2^{-\frac{1}{2}}\frac{{{{\tilde I}_1}}}{{{I_1}}}{E_1} = {E_1}\\
{{\tilde E}_2} =  2^{-1}\left( {\frac{{{{\tilde I}_2}}}{{{I_2}}}{E_2} - 2\frac{{{{\tilde I}_2}}}{{{I_0}}}{E_0}} \right)\\
{{\tilde E}_3} =  2^{-\frac{3}{2}}\left( {\frac{{{{\tilde I}_3}}}{{{I_3}}}{E_2} - 6\frac{{{{\tilde I}_3}}}{{{I_1}}}{E_1}} \right)\\
{{\tilde E}_4} =  2^{-2}\left( {\frac{{{{\tilde I}_4}}}{{{I_4}}}{E_4} - 12\frac{{{{\tilde I}_4}}}{{{I_2}}}{E_2} + 12\frac{{{{\tilde I}_4}}}{{{I_0}}}{E_0}} \right),
\end{array}
\label{1.37}
\end{equation}
where the previously dropped factor should be considered as included on both sides.\\
Comparing Eq.~\eqref{1.37} with the Hermite polynomials, Table~\ref{Tab_1},  reveals that the numerical coefficients in Eq.~\eqref{1.37} follow the coefficients of the Hermite polynomials.\\   
Without further proof it may hence be expected that the general solution can be written as
\begin{equation}
{\tilde E_n} =  {\left( { - 1} \right)^n}\sum\limits_{l + 2m = n} {\frac{{n!}}{{l!m!}}} {\left( { - 1} \right)^{l + m}}2^{-\frac{n}{2}}\frac{{{{\tilde I}_n}}}{{{I_l}}}{E_l},
\label{1.38}
\end{equation}
where the first part is given by the generating function of the Hermite polynomials (cf. Eq.~\eqref{1.3b}). Based on Eq.~\eqref{1.37} the sums ${S_{\text{in}}}$ and ${S_2}$, as well as $\beta$-function and emittance can be calculated (Eqs.~\eqref{1.22} and \ref{1.23}). The results are of course identical to the $\beta$-function and emittance given by Eqs.~\eqref{1.29} and \ref{1.30}.\\
The decomposition reveals a strong contribution of lower order modes to a field described by an higher order elegant Hermite-Gauss solution, which explains the relatively good beam quality of these fields even for high $n$.

\section{Conclusion}
The coherent superposition of modes leads to variations of characteristic beam parameters, which makes it necessary to distinguish the generating parameters and the observable parameters of the coherent sum. Nevertheless, the observable beam size of a coherent sum  follows in any case the standard rms envelope equation, and thus, the beam size can be described by a beam quality factor and a $\beta$-function, just as an incoherent beam. The rms envelope equation and the Courant-Snyder theory are hence established as general framework for the description of beams, which is suitable to tackle propagation, matching and imaging problems. This statements holds, whether the underlying decomposition of modes is known or not.\\
The generating parameters follow, of course, also the standard relations but with its own set of initial parameters. Modes, or coherent sums of modes, can thus be efficiently propagated through linear optical systems with simple methods, if the initial parameters are known. While the propagation of the generating functions allows the determination of the transverse field and intensity distribution, the observable parameters determine the imaging and matching conditions. The relations of the initial generating and observable parameters are derived for various cases in this report.\\
Finally, as an example for generalized solutions of the paraxial Helmholtz equation, the elegant Hermite-Gauss solution is interpreted as a coherent sum of standard Hermite-Gauss modes. While the $\beta$-function in case of the standard Hermite-Gauss modes is independent of the mode number (and equal to the generating $\beta$-function for pure modes) it decreases with increasing order of the solution in case of the elegant Hermite-Gauss solution. The decomposition reveals a strong contribution of lower order components in the higher order elegant solution, which explains the comparatively weak scaling of the beam quality with the mode number.

%%%%%%%%%%%%%%%%%%%%%%% References %%%%%%%%%%%%%%%%%%%%%%%%%

\end{document}